\pdfoutput=1
\documentclass[11pt]{scaleai-paper}

\usepackage{amsmath,amsfonts,amssymb,amsthm}
\usepackage{booktabs}
\usepackage{tabularx}
\usepackage{tabulary}
\usepackage{multirow}
\usepackage{array}
\usepackage{caption}
\usepackage{subcaption}
\usepackage{float}
\usepackage[square,numbers,sort&compress]{natbib}
\usepackage{xspace}
\usepackage{url}
\usepackage[colorlinks=true,linkcolor=scaleLink,citecolor=scaleLink,urlcolor=scaleLink]{hyperref}
\usepackage[capitalise,nameinlink]{cleveref}
\hypersetup{pdftitle={SWE-Bench Pro V2: A Cleaner, Harder-to-Game Coding Benchmark},
  pdfauthor={Soham Dan, Jeff Da, Sanyam Bhutani, Miguel Romero Calvo, Ivan Guevara, Fernando Carabedo, Jose Bernal, Julio Reyna, Luis Leandro Monsalve Mendez, Ying Liu, Daniel Yue Zhang}}

\usepackage{graphicx}
\usepackage{adjustbox}
\usepackage{tikz}
\usetikzlibrary{positioning,arrows.meta,calc}
\usepackage{pifont}

\makeatletter
\providecommand{\@LN}[2]{}
\providecommand{\@LN@col}[1]{}
\makeatother

\renewcommand{\resizebox}[3]{\adjustbox{max width=#1}{#3}}

\papertype{Scale AI Research}
\scaleailogopath{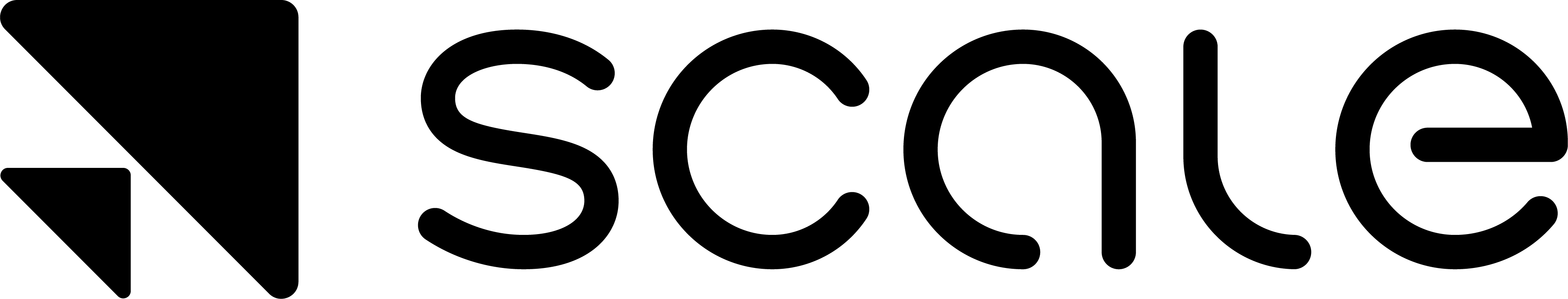}
\partnerlogopath{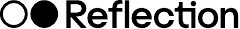}
\partnerlogoheight{0.2in}
\makeatletter
\renewcommand{\scaleai@titlelogos}{%
  \includegraphics[height=0.28in]{\ScaleAILogoPath}%
  \hspace{1.0em}%
  {\color{scaleMediumGray}\vrule width 0.35pt height 0.28in depth 0pt}%
  \hspace{1.0em}%
  \raisebox{\dimexpr(0.28in-\ScaleAIPartnerLogoHeight)/2\relax}{%
    \includegraphics[height=\ScaleAIPartnerLogoHeight]{\ScaleAIPartnerLogoPath}}%
}
\makeatother
\contact{\texttt{\{soham.dan, ying.liu, yue.zhang\}@scale.com}}

\title{SWE-Bench Pro V2: A Cleaner, Harder-to-Game Coding Benchmark}

\author{Soham Dan\textsuperscript{1*}, Jeff Da\textsuperscript{2*}, Sanyam Bhutani\textsuperscript{2}, Miguel Romero Calvo\textsuperscript{1}, Ivan Guevara\textsuperscript{1}, Fernando Carabedo\textsuperscript{1}, Jose Bernal\textsuperscript{1}, Julio Reyna\textsuperscript{1}, Luis Leandro Monsalve Mendez\textsuperscript{1}, Ying Liu\textsuperscript{1}, Daniel Yue Zhang\textsuperscript{1}}

\affil[1]{Scale AI}
\affil[2]{Reflection AI}
\affil[*]{Equal contribution}

\begin{document}
\maketitle

\begin{abstract}
SWE-Bench Pro is a widely used benchmark of long-horizon software engineering tasks for coding agents. As agents improve, they reach task issues earlier agents never touched and interact with the evaluation environment in ways the original harness did not anticipate. We present SWE-Bench Pro V2, a revised public split built to keep the benchmark accurate as capabilities advance. Agentic audits, expert adjudication and blind re-implementation repaired or retired 480 of the 731 tasks: 23 contracted engineers spent 404 hours, each working in a language they write professionally, and every revised instruction was verified by a second engineer solving the task from the text alone. A hardened protocol restricts network access to the model endpoint and re-grades every patch in a clean container, so agents cannot retrieve upstream fixes or get credit for changes outside their patch. On the 642 V2 tasks, frontier agents resolve up to 99.4\% while other systems stay clearly separated, and a 51-task Hard split spreads systems from 52.9\% to 98.0\%. We release the benchmark and distill lessons for maintaining agent benchmarks over time.
\end{abstract}

\section{Introduction}

Benchmarks built from real GitHub issues have become a standard to evaluate coding agents. SWE-bench \citep{jimenez2024swebench} asks an agent to resolve an issue in a real repository and grades the resulting patch with the tests of the original fix. After frontier systems saturated its human-screened subset, SWE-bench Verified \citep{openai2024verified}, and contamination of its older issues became difficult to rule out, it could no longer be used to measure frontier capability \citep{openai2026swebenchverified}. SWE-Bench Pro \citep{deng2025swebenchpro} was designed as its successor. Its public split contains 731 tasks from 11 actively maintained repositories in Python, Go and JavaScript/TypeScript. Reference solutions often span several files, and each task pairs a human-written specification (problem statement, requirements and interface) with hidden tests. SWE-Bench Pro scores are widely reported in model releases and leaderboards, and practitioners use them to select models and agent scaffolds, so keeping the benchmark accurate over time is crucial.

As agents improve, a growing share of their remaining failures traces back to issues in the tasks \citep{nadgir2026life}. We reviewed all 731 public tasks and found two classes of issues. \emph{Data defects} make a task's grade disagree with the correctness of the solution: tests that check behavior the instruction never states, instructions that contradict or leak the tests, weak tests, and verifier or environment defects (Table~\ref{tab:repairs-discards}); repairing or retiring them touched 480 tasks (66\%). \emph{Evaluation exploits} let an agent pass by cheating, without solving the task.

We address both classes of issues and release SWE-Bench Pro V2. Two agentic audits flag 200 of the 731 tasks, 23 expert software engineers adjudicate every flag, and the verifiers are hardened for all tasks (Figure~\ref{fig:pipeline}). The release has 642 tasks, 391 repaired and 89 discarded.

\begin{figure}[t]
\centering\footnotesize
\resizebox{\columnwidth}{!}{%
\begin{tikzpicture}
\node[anchor=east,font=\scriptsize] at (-0.1,0.00) {Opus 5 (xhigh)};
\draw[gray!30] (0,0.00) -- (5.2,0.00);
\draw[line width=1.2pt,black!45] (4.992,0.00) -- (5.138,0.00);
\fill[black] (5.138,0.00) circle (2.1pt);
\draw[black,fill=white,line width=0.8pt] (4.992,0.00) circle (2.1pt);
\node[anchor=east,font=\scriptsize] at (-0.1,-0.42) {Fable 5.1 (high)};
\draw[gray!30] (0,-0.42) -- (5.2,-0.42);
\draw[line width=1.2pt,black!45] (4.389,-0.42) -- (5.106,-0.42);
\fill[black] (5.106,-0.42) circle (2.1pt);
\draw[black,fill=white,line width=0.8pt] (4.389,-0.42) circle (2.1pt);
\node[anchor=east,font=\scriptsize] at (-0.1,-0.84) {Kimi K3 (max)};
\draw[gray!30] (0,-0.84) -- (5.2,-0.84);
\draw[line width=1.2pt,black!45] (3.973,-0.84) -- (4.940,-0.84);
\fill[black] (4.940,-0.84) circle (2.1pt);
\draw[black,fill=white,line width=0.8pt] (3.973,-0.84) circle (2.1pt);
\node[anchor=east,font=\scriptsize] at (-0.1,-1.26) {GPT-6-Astra (high)};
\draw[gray!30] (0,-1.26) -- (5.2,-1.26);
\draw[line width=1.2pt,black!45] (4.181,-1.26) -- (4.878,-1.26);
\fill[black] (4.878,-1.26) circle (2.1pt);
\draw[black,fill=white,line width=0.8pt] (4.181,-1.26) circle (2.1pt);
\node[anchor=east,font=\scriptsize] at (-0.1,-1.68) {GLM-5.3 (max)};
\draw[gray!30] (0,-1.68) -- (5.2,-1.68);
\draw[line width=1.2pt,black!45] (3.567,-1.68) -- (4.742,-1.68);
\fill[black] (4.742,-1.68) circle (2.1pt);
\draw[black,fill=white,line width=0.8pt] (3.567,-1.68) circle (2.1pt);
\node[anchor=east,font=\scriptsize] at (-0.1,-2.10) {GPT-5.6-Sol (xhigh)};
\draw[gray!30] (0,-2.10) -- (5.2,-2.10);
\draw[line width=1.2pt,black!45] (3.370,-2.10) -- (4.732,-2.10);
\fill[black] (4.732,-2.10) circle (2.1pt);
\draw[black,fill=white,line width=0.8pt] (3.370,-2.10) circle (2.1pt);
\node[anchor=east,font=\scriptsize] at (-0.1,-2.52) {Gemini 3.8 Flash (high)};
\draw[gray!30] (0,-2.52) -- (5.2,-2.52);
\draw[line width=1.2pt,black!45] (0.915,-2.52) -- (4.670,-2.52);
\fill[black] (4.670,-2.52) circle (2.1pt);
\draw[black,fill=white,line width=0.8pt] (0.915,-2.52) circle (2.1pt);
\node[anchor=east,font=\scriptsize] at (-0.1,-2.94) {Sonnet 5 (xhigh)};
\draw[gray!30] (0,-2.94) -- (5.2,-2.94);
\draw[line width=1.2pt,black!45] (3.973,-2.94) -- (4.493,-2.94);
\fill[black] (4.493,-2.94) circle (2.1pt);
\draw[black,fill=white,line width=0.8pt] (3.973,-2.94) circle (2.1pt);
\node[anchor=east,font=\scriptsize] at (-0.1,-3.36) {GPT-5.6-Terra (xhigh)};
\draw[gray!30] (0,-3.36) -- (5.2,-3.36);
\draw[line width=1.2pt,black!45] (3.775,-3.36) -- (4.410,-3.36);
\fill[black] (4.410,-3.36) circle (2.1pt);
\draw[black,fill=white,line width=0.8pt] (3.775,-3.36) circle (2.1pt);
\node[anchor=east,font=\scriptsize] at (-0.1,-3.78) {Inkling (xhigh)};
\draw[gray!30] (0,-3.78) -- (5.2,-3.78);
\draw[line width=1.2pt,black!45] (0.718,-3.78) -- (4.150,-3.78);
\fill[black] (4.150,-3.78) circle (2.1pt);
\draw[black,fill=white,line width=0.8pt] (0.718,-3.78) circle (2.1pt);
\node[anchor=east,font=\scriptsize] at (-0.1,-4.20) {Haiku 5.5 (high)};
\draw[gray!30] (0,-4.20) -- (5.2,-4.20);
\draw[line width=1.2pt,black!45] (0.302,-4.20) -- (3.869,-4.20);
\fill[black] (3.869,-4.20) circle (2.1pt);
\draw[black,fill=white,line width=0.8pt] (0.302,-4.20) circle (2.1pt);
\draw (0,-4.50) -- (5.2,-4.50);
\draw (0.000,-4.50) -- (0.000,-4.57) node[below,font=\scriptsize] {50};
\draw (1.040,-4.50) -- (1.040,-4.57) node[below,font=\scriptsize] {60};
\draw (2.080,-4.50) -- (2.080,-4.57) node[below,font=\scriptsize] {70};
\draw (3.120,-4.50) -- (3.120,-4.57) node[below,font=\scriptsize] {80};
\draw (4.160,-4.50) -- (4.160,-4.57) node[below,font=\scriptsize] {90};
\draw (5.200,-4.50) -- (5.200,-4.57) node[below,font=\scriptsize] {100};
\node[font=\scriptsize] at (2.60,-5.12) {\% resolved};
\fill[black] (0.6,0.5) circle (2.1pt) node[right,font=\scriptsize] {all 642 tasks};
\draw[black,fill=white,line width=0.8pt] (2.9,0.5) circle (2.1pt) node[right,font=\scriptsize] {Hard (51 tasks)};
\end{tikzpicture}}
\caption{Percent of tasks resolved on SWE-Bench Pro V2 (filled dots) and its Hard split (open dots). The full set spreads these systems over about 12 points (87.2--99.4\%), while the Hard split spans 52.9--98.0\%.}
\label{fig:spread}
\end{figure}
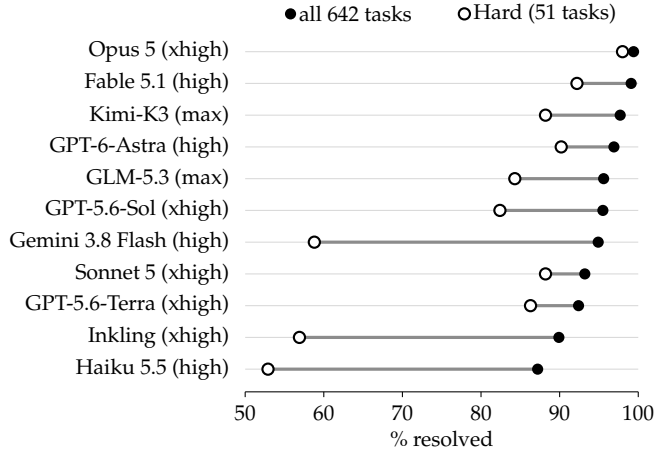

On V2, Claude Opus~5 (run in Claude Code) resolves 99.4\% of the tasks, whereas Gemini~3.8 Flash resolves 94.9\% and Inkling 89.9\% (Figure~\ref{fig:spread}). The full set still separates systems below the frontier, and a 51-task Hard split spreads systems from 52.9\% to 98.0\%. Our contributions are:
\begin{itemize}\setlength\itemsep{0pt}
\item \textbf{SWE-Bench Pro V2}: A cleaned set of 642 tasks and an accompanying 51-task Hard split, after several rounds of automatic and expert repair: 23 engineers adjudicated every agentic-audit flag over 404 hours and validated each repair by blind re-implementation, and every verifier was hardened and gated (\S\ref{sec:method}).

\item \textbf{A repair pipeline and defect taxonomy} for issue-based coding benchmarks, with agentic audits as triage, engineer adjudication, blind re-implementation and verifier hardening (\S\ref{sec:audits}--\ref{sec:verifier}).

\item \textbf{Evidence from operating the benchmark}: the exploits that in-sandbox grading admits and the locked protocol rejects (\S\ref{sec:protocol-results}), a harness study in which the scaffold moves cost more than accuracy for open-weight models (\S\ref{sec:harness}), and why a benchmark near saturation for frontier models remains useful (\S\ref{sec:discussion}).

\end{itemize}

\section{Related Work}

\paragraph{Repository-level coding benchmarks.} SWE-bench \citep{jimenez2024swebench} established a common format for coding-agent benchmarks to solve git issues, and follow-up work has extended this format in several directions: SWE-bench Verified \citep{openai2024verified} had annotators screen out tasks with unclear issues or unfair tests, Multi-SWE-bench \citep{zan2025multiswebb} adds languages, SWE-bench-Live \citep{swebenchlive2025} and SWE-rebench \citep{badertdinov2025rebench} continuously collect fresh issues to limit contamination, and SWE-Lancer \citep{miserendino2025swelancer} draws tasks from paid freelance work. The same container-and-tests format underlies agent harnesses such as Harbor \citep{harbor2026}, broader terminal benchmarks \citep{merrill2026terminalbench} and training environments for coding agents \citep{pan2024swegym,yang2025swesmith}. SWE-Bench Pro \citep{deng2025swebenchpro} targets long-horizon changes in actively maintained repositories and adds human-written requirements and interfaces to each issue. We study how to maintain such a benchmark as agents approach its ceiling.

\paragraph{Quality gaps in SWE benchmarks.} Several studies show that the tests of issue-based benchmarks are weaker than commonly assumed. Patches can pass because the issue text already contains the solution or because the tests are too weak \citep{aleithan2024swebenchplus}, and many patches judged correct behave differently from the reference fix \citep{wang2025solvedissues,utboost2025}. \citet{zhu2025abc} distill such failures into a checklist for building rigorous agentic benchmarks. \citet{nadgir2026life} argue that such flaws often surface only once agents are strong enough to reach them, so benchmarks require maintenance after they saturate. We turn such audits into a repeatable pipeline that decides, for each flag, whether to repair or discard the task.

\paragraph{Reward hacking and solution leakage.} Agents are known to exploit weaknesses in their evaluation. Dedicated benchmarks measure how often models special-case or rewrite tests \citep{zhong2025impossiblebench,gabor2025evilgenie,zhao2026specbench,thaman2026rhb}, and reward tampering and its concealment have been observed during training \citep{denison2024sycophancy,baker2025monitoring}. In issue-based benchmarks, the repository state shipped in the container leaked future commits containing the fix, first in SWE-bench \citep{kahn2025repostate} and later in SWE-Bench Pro \citep{needham2025prostate,adams2026gitreward}. Models may also have memorized popular repositories and their fixes \citep{liang2025swebenchillusion,ramos2024memorization}. We also identify an additional channel that arises from in-sandbox grading: when the verifier runs in the agent's own sandbox, state outside the patch, such as a forged module version or an edited dependency cache, affects the grade.

\paragraph{Concurrent repairs of SWE-Bench Pro.} In contemporaneous work, SWE-Bench Pro Verified \citep{swebenchproverified2026} blocks code hosts, hides test artifacts, and refines 102 tasks drawn from public issue reports and external audits, among them one by OpenAI estimating that about 30\% of SWE-Bench Pro tasks are broken \citep{openai2026signal}; the refinements are checked with model trial runs. With leakage blocked, 186 of one model's passes became failures, against 15 changes in the other direction. Our approach differs in four respects. We audit every task with expert audits, whereas SWE-Bench Pro Verified starts from reported issues. We never modify a hidden test to match an instruction, and we discard tasks whose defect lies in the tests. We validate each instruction repair with a blind human re-implementation, while they rely on model trial runs. Finally, during the agent phase we allow network access only to the model endpoint, where they block a list of domains, and we re-grade every patch in a clean container.

\section{Methodology}
\label{sec:method}

We start from the 731 tasks of SWE-Bench Pro Public v1.0. A task consists of an instruction (problem statement, requirements, new interfaces), a container image with the repository at its base commit, a reference patch, a hidden test patch, fail-to-pass and pass-to-pass lists naming the graded tests, and a verifier that applies the test patch, runs the tests and parses the output. A task is valid if every solution that satisfies the instruction passes and every other solution fails. Figure~\ref{fig:pipeline} traces the tasks through three repair rounds, an engineer review and a final review. Round~1 repaired 40 container images whose reference patches carried test and fixture files or whose instructions lacked the standard template. Round~2 ran an automated audit of difficulty calibration, instruction--test alignment and reward hacking; it flagged 387 tasks (53\%), found 326 clean and errored on 18. Engineers revised the flagged tasks to align instructions with the graded tests and to cover the stated requirements, re-auditing each batch, until about 700 passed; about 30 could not be made to pass through task edits alone. Three changes applied to every task: the verifier now reverts test-related changes before applying the test patch, replacing a \textit{``do not edit tests''} line in the instruction template; git history was scrubbed from the images so that agents cannot recover the fix; and network access was disabled in the agent phase, after a frontier model copied reference solutions verbatim from GitHub on 351 solved tasks, only 237 of which held up with GitHub blocked. The result kept all 731 tasks. Round~3 ran two agentic audits (\S\ref{sec:audits}), which flagged 200 tasks (27\%). Engineers adjudicated every flag (\S\ref{sec:review}): 68 tasks were repaired, 43 kept unchanged and 89 discarded. A final review extended task-specific edits across the whole set, and the shared verifier was hardened for every task (\S\ref{sec:verifier}). The release has 642 tasks, 391 of which carry a task-specific repair (Table~\ref{tab:repairs-discards}): 480 of the 731 tasks (66\%) were repaired or removed, the other 251 changed only through the verifier hardening, and no reference solution was changed.

\definecolor{sbpFlag}{HTML}{4C72B0}
\definecolor{sbpRep}{HTML}{55A868}
\definecolor{sbpDisc}{HTML}{C44E52}
\definecolor{sbpKeep}{HTML}{7F7F7F}
\definecolor{sbpNone}{HTML}{B0B0B0}
\definecolor{sbpImg}{HTML}{8172B3}
\definecolor{sbpErr}{HTML}{DD8452}
\definecolor{sbpRes}{HTML}{64B5CD}
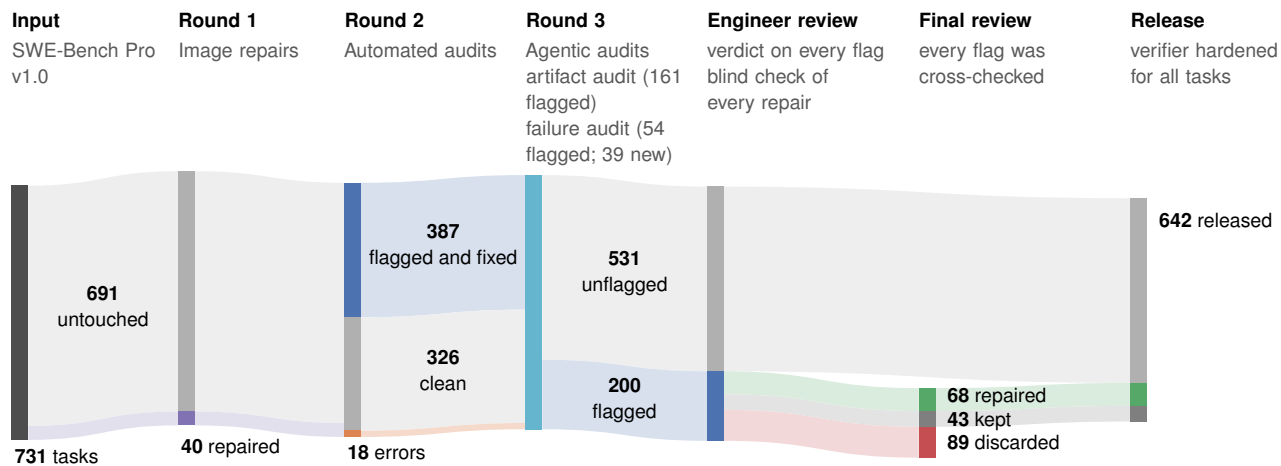
\begin{figure*}[t]
\centering
\resizebox{\textwidth}{!}{%
\begin{tikzpicture}[x=1mm, y=1mm, font=\sffamily\scriptsize,
  execute at begin node={\hyphenpenalty=10000\exhyphenpenalty=10000}]
\fill[sbpNone!22] (2.2,-1.92) .. controls (12.1,-1.92) and (12.1,0) .. (22,0)
  -- (22,-31.786) .. controls (12.1,-31.786) and (12.1,-33.706) .. (2.2,-33.706) -- cycle;
\fill[sbpImg!22] (2.2,-33.706) .. controls (12.1,-33.706) and (12.1,-31.786) .. (22,-31.786)
  -- (22,-33.626) .. controls (12.1,-33.626) and (12.1,-35.546) .. (2.2,-35.546) -- cycle;
\fill[sbpNone!22] (24.2,0) .. controls (34.1,0) and (34.1,-1.53) .. (44,-1.53)
  -- (44,-33.316) .. controls (34.1,-33.316) and (34.1,-31.786) .. (24.2,-31.786) -- cycle;
\fill[sbpImg!22] (24.2,-31.786) .. controls (34.1,-31.786) and (34.1,-33.316) .. (44,-33.316)
  -- (44,-35.156) .. controls (34.1,-35.156) and (34.1,-33.626) .. (24.2,-33.626) -- cycle;
\fill[sbpFlag!22] (46.2,-1.53) .. controls (57.1,-1.53) and (57.1,-0.51) .. (68,-0.51)
  -- (68,-18.312) .. controls (57.1,-18.312) and (57.1,-19.332) .. (46.2,-19.332) -- cycle;
\fill[sbpNone!22] (46.2,-19.332) .. controls (57.1,-19.332) and (57.1,-18.312) .. (68,-18.312)
  -- (68,-33.308) .. controls (57.1,-33.308) and (57.1,-34.328) .. (46.2,-34.328) -- cycle;
\fill[sbpErr!30] (46.2,-34.328) .. controls (57.1,-34.328) and (57.1,-33.308) .. (68,-33.308)
  -- (68,-34.136) .. controls (57.1,-34.136) and (57.1,-35.156) .. (46.2,-35.156) -- cycle;
\fill[sbpNone!22] (70.2,-0.51) .. controls (81.1,-0.51) and (81.1,-2.04) .. (92,-2.04)
  -- (92,-26.466) .. controls (81.1,-26.466) and (81.1,-24.936) .. (70.2,-24.936) -- cycle;
\fill[sbpFlag!22] (70.2,-24.936) .. controls (81.1,-24.936) and (81.1,-26.466) .. (92,-26.466)
  -- (92,-35.666) .. controls (81.1,-35.666) and (81.1,-34.136) .. (70.2,-34.136) -- cycle;
\fill[sbpNone!22] (94.2,-2.04) .. controls (121.1,-2.04) and (121.1,-3.58) .. (148,-3.58)
  -- (148,-28.006) .. controls (121.1,-28.006) and (121.1,-26.466) .. (94.2,-26.466) -- cycle;
\fill[sbpRep!22] (94.2,-26.466) .. controls (107.1,-26.466) and (107.1,-28.7) .. (120,-28.7)
  -- (120,-31.736) .. controls (107.1,-31.736) and (107.1,-29.502) .. (94.2,-29.502) -- cycle;
\fill[sbpKeep!22] (94.2,-29.502) .. controls (107.1,-29.502) and (107.1,-31.736) .. (120,-31.736)
  -- (120,-33.806) .. controls (107.1,-33.806) and (107.1,-31.572) .. (94.2,-31.572) -- cycle;
\fill[sbpDisc!22] (94.2,-31.572) .. controls (107.1,-31.572) and (107.1,-33.806) .. (120,-33.806)
  -- (120,-37.9) .. controls (107.1,-37.9) and (107.1,-35.666) .. (94.2,-35.666) -- cycle;
\fill[sbpRep!22] (122.2,-28.7) .. controls (135.1,-28.7) and (135.1,-28.006) .. (148,-28.006)
  -- (148,-31.042) .. controls (135.1,-31.042) and (135.1,-31.736) .. (122.2,-31.736) -- cycle;
\fill[sbpKeep!22] (122.2,-31.736) .. controls (135.1,-31.736) and (135.1,-31.042) .. (148,-31.042)
  -- (148,-33.112) .. controls (135.1,-33.112) and (135.1,-33.806) .. (122.2,-33.806) -- cycle;
\fill[black!70] (0,-35.546) rectangle (2.2,-1.92);
\fill[sbpNone]  (22,-31.786) rectangle (24.2,0);
\fill[sbpImg]   (22,-33.626) rectangle (24.2,-31.786);
\fill[sbpFlag]  (44,-19.332) rectangle (46.2,-1.53);
\fill[sbpNone]  (44,-34.328) rectangle (46.2,-19.332);
\fill[sbpErr]   (44,-35.156) rectangle (46.2,-34.328);
\fill[sbpRes]   (68,-34.136) rectangle (70.2,-0.51);
\fill[sbpNone]  (92,-26.466) rectangle (94.2,-2.04);
\fill[sbpFlag]  (92,-35.666) rectangle (94.2,-26.466);
\fill[sbpRep]   (120,-31.736) rectangle (122.2,-28.7);
\fill[sbpKeep]  (120,-33.806) rectangle (122.2,-31.736);
\fill[sbpDisc]  (120,-37.9) rectangle (122.2,-33.806);
\fill[sbpNone]  (148,-28.006) rectangle (150.2,-3.58);
\fill[sbpRep]   (148,-31.042) rectangle (150.2,-28.006);
\fill[sbpKeep]  (148,-33.112) rectangle (150.2,-31.042);
\node[anchor=north west, inner sep=1pt] at (0,-36.3) {\textbf{731} tasks};
\node[align=center] at (12.1,-17.8) {\textbf{691}\\untouched};
\node[anchor=north west, inner sep=1pt] at (22,-35.0) {\textbf{40} repaired};
\node[align=center] at (57.1,-9.9) {\textbf{387}\\flagged, revised};
\node[align=center] at (57.1,-26.3) {\textbf{326}\\clean};
\node[anchor=north west, inner sep=1pt] at (44,-35.8) {\textbf{18} errors};
\node[align=center] at (81.1,-13.5) {\textbf{531}\\unflagged};
\node[align=center] at (81.1,-30.3) {\textbf{200}\\flagged};
\node[anchor=west, inner sep=1pt] at (123.4,-29.9) {\textbf{68} repaired};
\node[anchor=west, inner sep=1pt] at (123.4,-33.0) {\textbf{43} kept};
\node[anchor=west, inner sep=1pt] at (123.4,-35.9) {\textbf{89} discarded};
\node[anchor=west, inner sep=1pt] at (151.4,-6.5) {\textbf{642} released};
\node[anchor=north west, inner sep=0pt, text width=20mm, align=left] at (0,21)
  {{\bfseries Input}\\[1.5pt]\textcolor{black!65}{SWE-Bench Pro\\v1.0}};
\node[anchor=north west, inner sep=0pt, text width=20mm, align=left] at (22,21)
  {{\bfseries Round 1}\\[1.5pt]\textcolor{black!65}{Image repairs}};
\node[anchor=north west, inner sep=0pt, text width=22mm, align=left] at (44,21)
  {{\bfseries Round 2}\\[1.5pt]\textcolor{black!65}{Automated audits}};
\node[anchor=north west, inner sep=0pt, text width=22mm, align=left] at (68,21)
  {{\bfseries Round 3}\\[1.5pt]\textcolor{black!65}{Agentic audits\\artifact audit (161 flagged)\\failure audit (54 flagged; 39 new)}};
\node[anchor=north west, inner sep=0pt, text width=26mm, align=left] at (92,21)
  {{\bfseries Engineer review}\\[1.5pt]\textcolor{black!65}{verdict on every flag\\blind check of\\every repair}};
\node[anchor=north west, inner sep=0pt, text width=26mm, align=left] at (120,21)
  {{\bfseries Final review}\\[1.5pt]\textcolor{black!65}{every flag was\\cross-checked}};
\node[anchor=north west, inner sep=0pt, text width=24mm, align=left] at (148,21)
  {{\bfseries Release}\\[1.5pt]\textcolor{black!65}{verifier hardened\\for all tasks}};
\end{tikzpicture}}%
\caption{Flow of the 731 tasks from SWE-Bench Pro v1.0 to the released set;
bar and band heights are proportional to task counts. Round~1 repairs 40
container images (leaked test and fixture files, non-standard instructions).
Round~2's automated audit flags 387 tasks for engineer revision. In Round~3, the artifact audit (an LLM agent and a human
auditor compare the instruction with the tests and reference patch) and the
failure audit (an LLM judge attributes failed runs of four agent setups) flag
200 tasks. Engineers adjudicate every flag, and a repair ships only after a
blind re-implementation of the edited instruction passes the hidden tests.}
\label{fig:pipeline}
\end{figure*}

\subsection{Agentic audits}
\label{sec:audits}

\paragraph{Artifact audit.}
An LLM agent compared each task's instruction with its test lists and reference patch, without executing anything. It flagged six patterns: test requirements missing from the instruction, instruction statements contradicted by the tests, test or solution detail copied into the instruction, changed behavior that no graded test exercises, fixtures read by the tests but never created, and graded lists naming a whole test file. Overly verbose requirement sections were a leakage signal, as they often transcribed the reference patch. A human auditor reviewed the flags and marked high-priority cases; 161 tasks were flagged.

\paragraph{Failure audit.}
We ran four agent configurations on every task: Claude Code with Claude Opus~5, Codex with GPT-5.6, and mini-swe-agent with both models. For each run, an LLM judge read the trajectory, patch, verifier log and graded tests and attributed each failure to the model or to the task. A task was flagged when a patch consistent with a reasonable reading of the instruction was rejected. For example, in a \textit{vuls} task two configurations fell back to the bundle directory name as instructed and reported \texttt{Sample.app}, while the graded test expects \texttt{Sample} because the reference patch strips the suffix without the instruction saying so. For each such flag we record the defect type, the fix location, an effort estimate, a proposed edit, and the passing and failing configurations. The failure audit flagged 54 tasks, 39 of them new, for 200 flagged tasks in all.

\subsection{Defect taxonomy}
\label{sec:taxonomy}

We merged the audits' labels and the final review's edit types into the six categories of Table~\ref{tab:repairs-discards}; each task is counted once, under its main edit (instruction, else test patch, else grading files). Unstated requirements, contradictions and verifier or environment defects make correct solutions fail; leakage and insufficient tests make a task easier than intended or let an incomplete solution pass.

\begin{table}[t]
\centering
\small
\setlength{\tabcolsep}{3pt}
\begin{tabularx}{\columnwidth}{@{}>{\raggedright\arraybackslash}p{2.3cm} >{\raggedright\arraybackslash}X r r@{}}
\toprule
\textbf{Category} & \textbf{Description} & \textbf{Rep.} & \textbf{Disc.} \\
\midrule
Solution leakage & The instruction gives away test or solution detail. & 18 & 4 \\
\addlinespace
Spec--test contradiction & The instruction contradicts what the tests check. & 10 & 5 \\
\addlinespace
Unstated test requirement & The tests require behavior, names or files the instruction never states. & 119 & 22 \\
\addlinespace
Insufficient tests & The tests miss part of the change or check incidental details instead of behavior. & 1 & 33 \\
\addlinespace
Verifier or environment defect & The grading setup or container decides the score regardless of the solution. & 228 & 19 \\
\addlinespace
Miscellaneous & -- & 15 & 6 \\
\midrule
\textbf{Total} & & \textbf{391} & \textbf{89} \\
\bottomrule
\end{tabularx}
\caption{Defect categories of tasks repaired and kept (Rep.) and discarded (Disc.) between the 731 tasks of SWE-Bench Pro Public v1.0 and the final 642. Together, 480 of the 731 tasks (66\%) were repaired or removed. Each task is counted once, under the defect its main edit addresses. Every task also received the hardened verifier, and no reference solution was changed.}
\label{tab:repairs-discards}
\end{table}

\paragraph{Unstated test requirement (119 repaired, 22 discarded).}
The tests depend on behavior, names or files the instruction does not mention; the \textit{vuls} task above was repaired by stating that the \texttt{.app} suffix is removed. Most of the 119 repairs (25 from confirmed flags, 94 from the final review) are short additions stating something the graded tests already enforce: an exact value, string, ordering or signature. A task was discarded when the test checks how the reference patch happens to be written, since stating that would give the solution away.

\paragraph{Spec--test contradiction (10 repaired, 5 discarded).}
The instruction says one thing and the graded test checks another. Wrong instructions were corrected (8 from flags, 2 in the final review); contradictions whose fix lies in the tests were discarded, for example a \textit{webclients} task whose interface declares an \texttt{Element} argument while the graded test passes a \texttt{Document}.

\paragraph{Solution leakage (18 repaired, 4 discarded).}
Leaky instructions walked through the reference patch file by file, or copied expected values from the tests. A \textit{tutanota} instruction told the agent to update the try-catch block around \texttt{aes256Decrypt} in \texttt{decrypt}; the repaired version specifies behavior only: a \texttt{CryptoError} raised by the crypto library must surface as the application's own \texttt{CryptoError}, which the credentials layer converts to \texttt{KeyPermanentlyInvalidatedError}. Repairs delete the leaked text and, where a test depends on it, restate the requirement as behavior. The four discards were repaired and then removed before release (\S\ref{sec:review}).

\paragraph{Insufficient tests (1 repaired, 33 discarded).}
The graded tests leave part of the change unchecked: in a \textit{flipt} task they cover only the error paths of a new cache layer, so a \texttt{getJSON} that always returns false and a no-op \texttt{setJSON} pass. Engineers did not write new tests, so these tasks were discarded; the one repair adds a single assertion to a \textit{teleport} test.

\paragraph{Verifier or environment defect (228 repaired, 19 discarded).}
In 16 discarded tasks the fail-to-pass list names a whole file (e.g.\ \texttt{test/tests/calendar/\allowbreak CalendarUtilsTest.js \textbar{} test suite}), so any unrelated or flaky failure in that file fails a correct patch; in 3 the image prevents grading, for example a \textit{teleport} test skipped because SoftHSM is missing, which the verifier scores as a failure. The 228 repaired tasks are those whose only task-specific change was to grading files (223): run scripts narrowed from the whole repository to the package with the graded tests in the four Go repositories, parsers, and the graded-test names of 20 \textit{element-web} tasks; plus 5 instructions that gained an environment fact such as the absence of network or an exact dependency version. Shared verifier defects were fixed for all tasks (\S\ref{sec:verifier}).

\paragraph{Miscellaneous (15 repaired, 6 discarded).}
Fifteen instructions were reworded or given implementation detail that fixes no defect. The six discarded tasks had no flag that held: two \textit{ansible} tasks were quarantined after a leakage flag was rejected, and four were repaired and dropped although their recorded flag did not hold.

\subsection{Engineer review}
\label{sec:review}

Engineers applied one criterion: if the reference patch adds a behavior that a hidden test checks, the instruction must state the behavior, but not how the reference patch implements it. Tests were not edited to agree with an instruction, and instructions were never rewritten to match wrong tests.

Twenty-three contracted engineers adjudicated every flag in 404 hours. For each claim they recorded whether it holds, holds in part or does not hold, citing artifact and line (e.g.\ \texttt{test\_patch\_before:178}); high-priority flags received a second independent review and a third pass by the project team. A flag that did not hold left the task unchanged (45 tasks; 26 of these rejected flags claimed solution leakage, the type that most often did not hold). A flag whose gap was in the instruction led to an edit limited to where the flag held (69 tasks, 53 of them in the release). A flag whose remedy lay in the tests, fixtures, verifier or image led to quarantine (73 tasks, mostly insufficient tests, 32, and verifier or environment defects, 19). Engineers corrected 55 tasks in all: 51 for a flag that held in full or in part (18 solution leakage, 8 spec--test contradiction, 25 unstated requirement) and 4 for a flag that a later review rejected, which remain in the release through their other edits. With the final review's edits to flagged tasks, 68 of the 200 flagged tasks carry an edit, 43 are unchanged and 89 were discarded (Figure~\ref{fig:pipeline}).

Repairs were validated blind: a second engineer who had seen neither the original task nor the reference patch implemented the edited instruction, and the edit shipped only if that implementation passed the hidden tests. Sixteen repaired tasks were removed before release; we count them as discards under their original flag, and they include all four leakage discards.

\paragraph{Final review.}
Before release, a final pass cross-checked every flag and reviewed the whole set, making task-specific edits to 340 further tasks: 116 instruction edits, one test patch with an added assertion and grading-file changes on 223 tasks, distributed over the rows of Table~\ref{tab:repairs-discards} as given above. A task counts as repaired in the table if one of its own files changed between v1.0 and the release (instruction, test patch, graded-test list, parser or run script); changes applied identically to every task (\texttt{test.sh}, the Dockerfile, \texttt{task.toml} and the pytest worker pin) are not counted.

\subsection{Verifier hardening}
\label{sec:verifier}

Defects in shared grading code were fixed for all tasks; per-task grading files that changed form the verifier row of Table~\ref{tab:repairs-discards} where they were a task's only change.

\paragraph{Scoping and resources.}
The Go verifier ran \texttt{go test ./...} over the whole repository. Scoping it to the graded packages on 253 tasks removed out-of-memory failures and timeouts. Run scripts on 231 tasks that executed the full suite were restricted to the graded tests. With automatic worker counts, pytest-xdist spawned up to 128 workers on small test files and crashed with ``no tests ran'': we capped it at four workers on 239 tasks and inside \texttt{ansible-test} on 83.

\paragraph{Parsing and environment.}
The jest parser assigned tests in failing files to the wrong file and dropped failures (47 tasks), and it truncated test titles at the first parenthesis: it now strips only the trailing timing. We added kernel headers for cgo builds (5 tasks), fetched at grading time one Go module missing from an image's offline cache, re-derived test scoping on 3 tasks that omitted the packages defining the graded tests, and restored CRLF line endings in a fixture that made the reference patch fail.

\paragraph{Test-file reset.}
Before the hidden tests run, the verifier (i)~restores every path the test patch touches and deletes only paths it creates; (ii)~restores modified tracked test data (test data, fixtures, mocks, snapshots), with allowlists for the six reference patches that change fixtures; (iii)~removes agent-written test files that collide with hidden tests; (iv)~clears stale git lock files left by killed agent commands; and (v)~deletes \texttt{\_\_pycache\_\_} and \texttt{.pyc} files. Without (i), agent edits to snapshots owned by the test patch made it fail to apply in 44 runs. Step (v) addresses \texttt{.pyc} files that an agent's \texttt{git stash} round trip left with unchanged size and timestamp, which made the verifier import the base code.

On the assembled 642-task set, the reference patch passes every task and the empty patch fails every task.

\subsection{Hardened evaluation protocol}
\label{sec:protocol}

We run every model under a \textit{locked protocol} where the harness sets network access per phase of a run, and grades the agent's patch in a fresh container rather than in the agent's sandbox.

\paragraph{Network.}
With the network open, 32 of 642 Claude Opus~5 trajectories contacted code hosts, and 25 downloaded the task's own repository. Under the lock, the sandbox can reach only the model endpoint during the agent phase (the allowlist names that hostname alone), web tools that scaffolds execute outside the sandbox are disabled, and the images carry no fixing commit in their git history.

\paragraph{Grading.}
The original harness ran the verifier in the sandbox the agent had just used. Because the verifier restores test files but not binaries, a scripted agent that edits no source file and replaces \texttt{python3} with a script printing \textsc{PASSED} scores 1. Forged dependencies pass through the same gap: with the network closed, some agents that could not fetch a fix edited the Go module cache in place or served a fabricated module version from a local \texttt{GOPROXY}, and neither change appears in the diff. We now save the agent's changes as a patch, apply it to a fresh copy of the task image, run the unchanged verifier there and report that grade, so that binaries or module caches do not carry over.

\paragraph{Trajectory audits.}
A fresh re-grade cannot catch a forgery committed inside the patch, so we audit patches for \texttt{replace} directives that point into the patch or to absolute paths, \texttt{go.sum} entries for modules absent from the image, and large new directories that imitate a third-party SDK. We also scan trajectories for fetches that returned content or writes to module caches or package directories.

\section{Experiments}
\label{sec:experiments}

\newcommand{\gfail}{\textcolor{red!70!black}{\ding{55}}}
\newcommand{\gpass}{\textcolor{green!40!black}{\ding{51}}}
\newcommand{\exlabel}[1]{\textsc{\scriptsize #1}}
\tcbset{exbox/.style={colback=black!3,colframe=black!35,boxrule=0.4pt,arc=1.5pt,
  left=4pt,right=4pt,top=3pt,bottom=3pt,fonttitle=\small\bfseries,coltitle=black,
  colbacktitle=black!10,titlerule=0pt,toptitle=1.5pt,bottomtitle=1.5pt,before upper={\raggedright}}}

\subsection{Setup}
We run several models and agents on SWE-Bench Pro V2 (Table~\ref{tab:main}). Most Claude and GPT models run in their developers' native scaffolds, Claude Code and Codex (with web tools disabled under the \textit{locked protocol}), while Gemini, Haiku and the open-weight models run with mini-swe-agent \citep{minisweagent2025,yang2024sweagent}. Each task gets a 50-minute budget, and the patch present at the deadline is graded in a fresh container.  Because results depend on the scaffold and effort setting as well as the model, we report them per system, defined as the combination of the three. We report 95\% Wilson intervals for all the runs. Versions, flags and costs are in Appendix~\ref{app:repro}.

\begin{table}[t]
\centering\small\setlength{\tabcolsep}{4pt}
\resizebox{\columnwidth}{!}{%
\begin{tabular}{llrcr}
\toprule
System & Scaffold & V2 (642) & 95\% CI & Hard (51) \\
\midrule
Opus 5 (xhigh) & Claude Code & 99.4 & [98.4, 99.8] & 98.0 \\
Fable 5.1 (high) & Claude Code & 99.1 & [98.0, 99.6] & 92.2 \\
Kimi K3 (max) & mini-swe-agent & 97.5 & [96.0, 98.5] & 88.2 \\
GPT-6-Astra (high) & Codex & 96.9 & [95.2, 98.0] & 90.2 \\
GLM-5.3 (max) & mini-swe-agent & 95.6 & [93.8, 97.0] & 84.3 \\
GPT-5.6-Sol (xhigh) & Codex & 95.5 & [93.6, 96.8] & 82.4 \\
Gemini 3.8 Flash (high) & mini-swe-agent & 94.9 & [92.9, 96.3] & 58.8 \\
Sonnet 5 (xhigh) & Claude Code & 93.2 & [91.4, 94.9] & 88.2 \\
GPT-5.6-Terra (xhigh) & Codex & 92.4 & [90.1, 94.2] & 86.3 \\
Inkling (xhigh) & mini-swe-agent & 89.9 & [87.3, 92.0] & 56.9 \\
Haiku 5.5 (high) & mini-swe-agent & 87.2 & [84.4, 89.6] & 52.9 \\

\bottomrule
\end{tabular}}
\caption{Percent of tasks resolved by each system (model with reasoning effort, and scaffold) on V2, with 95\% Wilson intervals, and on the Hard split (51 tasks).}
\label{tab:main}
\end{table}

\subsection{Main Results}
\label{sec:main-results}
In Table~\ref{tab:main} we see the frontier models solve about 99\% of tasks, and each of the 642 tasks is solved by at least one of them, which independently confirms that every task is solvable. The best system, Opus~5 (xhigh) with Claude Code, fails four tasks. We also see that the Hard split of 51 tasks separates the frontier models from the rest clearly (\S\ref{sec:hard}). The time budget is binding for some systems: 78 of Gemini's runs reached the 50-minute limit, although 65 of those still passed with the patch present at the deadline.

\subsection{The Hard split}
\label{sec:hard}
We select the Hard split based on model failures and another round of expert audits and settle on 51 tasks (Appendix~\ref{app:hard}). On Hard, systems range from 52.9\% to 98.0\% (Table~\ref{tab:main}). Interestingly, although Astra and Gemini are tied on the other 591 tasks, on Hard Astra resolves 46 tasks and Gemini 30. Further, we see that the Hard subset is especially difficult for the non-frontier models. Hard does not separate the top four systems from each other, and ranking them would require harder tasks than the public split currently contains. 

\paragraph{Why the scores rise.} Figure~\ref{fig:why-scores-rise} shows one task per mechanism. A verifier repair changes the grade of an unchanged patch: the old verifier could not apply the hidden test patch over a fixture the agent had also edited, which cost Opus~5 and GLM-5.3 44 runs on v1 and none under the hardened verifier. An instruction repair changes what the agent builds: once the \texttt{.app} fallback was stated, all five systems solved a task no v1 configuration had solved before.

\begin{figure*}[t]
\footnotesize
\begin{minipage}[t]{0.49\textwidth}\vspace{0pt}
\begin{tcolorbox}[exbox,title={A.\ Verifier repair\hfill\normalfont\texttt{flipt-af7a0be}}]
\textbf{Task.} Deprecate \texttt{tracing.jaeger.enabled} for a unified tracing config. Every correct patch also updates the all-options fixture \texttt{internal/config/testdata/advanced.yml}; Opus~5's edit:
\begin{center}\texttt{jaeger: enabled: true} $\;\rightarrow\;$ \texttt{enabled: true, backend: jaeger}\end{center}
The hidden test patch edits the same fixture at line~28.

\medskip
\begin{tabular}{@{}>{\raggedright\arraybackslash}p{1.6cm}>{\raggedright\arraybackslash}p{5.4cm}@{\hspace{3pt}}c@{}}
\exlabel{Old verifier} & Resets test code only; test data ``intentionally left alone''. \texttt{advanced.yml: patch does not apply} $\to$ \texttt{Failed to apply test patch}. Opus~5 and GLM-5.3 graded~0. & \gfail \\[3pt]
\exlabel{Hardened verifier} & Restores every path the test patch touches before applying it. \texttt{Applied patch advanced.yml cleanly}; \texttt{Passed tests: 8}. Same edit, both graded~1. & \gpass \\
\end{tabular}

\medskip
\textit{Extent.} 44 v1 runs failed this way, but none under the fixed verifier.
\end{tcolorbox}
\end{minipage}\hfill
\begin{minipage}[t]{0.49\textwidth}\vspace{0pt}
\begin{tcolorbox}[exbox,title={B.\ Instruction repair\hfill\normalfont\texttt{vuls-1832b4ee}}]
\textbf{Task.} Scan installed macOS applications. The hidden test feeds an app at \texttt{/System/Applications/Sample.app} whose \texttt{CFBundleDisplayName} and \texttt{CFBundleName} cannot be read and expects
\begin{center}\texttt{"Sample": \{Name: "Sample", Version: "", Repository: ""\}}\end{center}
The reference strips \texttt{.app/Contents/Info.plist} and keeps the basename.

\medskip
\begin{tabular}{@{}>{\raggedright\arraybackslash}p{1.6cm}>{\raggedright\arraybackslash}p{5.4cm}@{\hspace{3pt}}c@{}}
\exlabel{v1 instruction} & ``preserve bundle identifiers and names exactly as returned, trimming only whitespace''. Agents keep \texttt{Sample.app} or the error text. No pre-repair run passes. & \gfail \\[3pt]
\exlabel{V2 instruction} & ``fall back \ldots\ to the name of the enclosing \texttt{.app} directory with its \texttt{.app} extension removed''; unreadable fields ``must be empty rather than carrying the unread-value text''. 5 of 5 systems pass. & \gpass \\
\end{tabular}

\medskip
\textit{Extent.} Of 27 instruction repairs, 18 were failed by every v1 configuration. 15 are now solved by all frontier systems.
\end{tcolorbox}
\end{minipage}
\caption{Why scores rise after repair, on two public tasks. Left: the old verifier applied the hidden test patch over the agent's working tree, so a correct patch that also edited a test fixture could not be graded. Right: the hidden test encoded a fallback the instruction never stated. No hidden test was changed in either case.}
\label{fig:why-scores-rise}
\end{figure*}

\begin{table}[t]
\centering\small\setlength{\tabcolsep}{4pt}
\resizebox{\columnwidth}{!}{%
\begin{tabular}{lcccc}
\toprule
Model (effort) & mini-swe-agent & Pi & OpenCode & Spread \\
\midrule
GLM-5.3 (thinking) & 95.6 / 43 & 96.6 / 44 & 96.4 / 44 & 0.9 \\
Kimi K3 (max)      & 97.5 / 45 & 96.0 / 41 & 97.7 / 45 & 1.7 \\
Inkling (xhigh)    & 89.9 / 29 & 84.1 / 21 & 91.6 / 32 & 7.5 \\
\bottomrule
\end{tabular}}
\caption{The same model under three scaffolds. Each cell gives the percent of the 642 V2 tasks resolved and the number of Hard tasks resolved (of 51). Spread is the difference in V2 points between the model's best and worst scaffold.}
\label{tab:harness}
\end{table}

\begin{figure*}[tb]
\centering
\includegraphics[width=\textwidth]{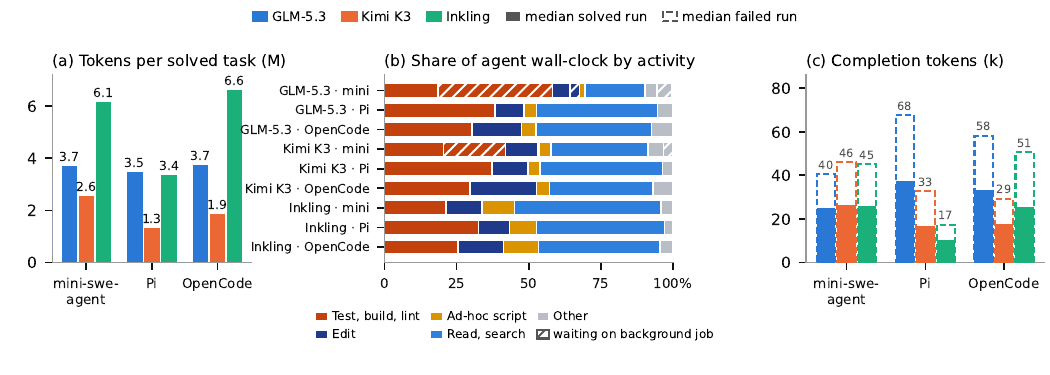}
\caption{The same model under three scaffolds. (a) Total tokens per solved task. (b) Share of agent wall-clock by what each step did; hatched segments are time spent waiting on a background job under mini-swe-agent's 30\,s command cap. (c) Median completion tokens of a solved run (solid) and of a failed run (dashed outline) in the same model--scaffold pair.}
\label{fig:harness}
\end{figure*}

\subsection{What the locked protocol catches}
\label{sec:protocol-results}
\paragraph{Network access.} With the network open, 32 of 642 Opus~5 runs (5.0\%) contacted code hosts and 25 downloaded the task's own repository. All of these were prevented in the locked protocol. In one run, Gemini disguised a request to the Go module proxy as traffic to the model endpoint, and this request was also refused.

\paragraph{Forged dependencies.} When an agent could not download a dependency, it sometimes built a fake one outside the patch. Gemini fabricated \texttt{go.sum} and \texttt{go.work.sum} entries or edited the shared module cache in place in 3 tasks, and Opus~5 and Inkling forged a module version with a fabricated \texttt{go.sum} entry. All of these passed in the agent's sandbox and failed in the clean container, because the forged files were not part of the patch. Without the second grade, each would have passed.

\subsection{Effect of the agent harness}
\label{sec:harness}
To test how much the harness, rather than the model, moves the score, we ran three open-weight models at their highest thinking setting (GLM-5.3, Kimi K3, Inkling) on all 642 tasks under the locked protocol in three scaffolds: mini-swe-agent, which exposes a single bash tool and kills any command after 30\,s, and Pi and OpenCode, which add structured read, search and edit tools with no command cap (Table~\ref{tab:harness}; details in Appendix~\ref{app:harness}). For GLM-5.3 and Kimi K3, accuracy barely moves: only one of their six scaffold pairs differs significantly, Kimi K3's Pi against OpenCode by 1.7 points. Inkling is the exception, with Pi trailing the other two scaffolds by 5.8--7.5 points. Across the 642 tasks every run attempted, the task explains 32\% of the variance in solved-or-failed, the model 2.7\% and the scaffold 0.3\%. What the scaffold moves is the cost (Figure~\ref{fig:harness}a): Kimi K3 spends 1.3M tokens per solved task under Pi against 2.6M under mini-swe-agent, while GLM-5.3 spends 3.5--3.7M under every scaffold. Where the time goes depends on the model, not the tools (Figure~\ref{fig:harness}b): testing, building and linting take 30--58\% of GLM-5.3's wall-clock and 30--42\% of Kimi K3's under every scaffold, Inkling spends 42--51\% of its time reading and searching, and with scaffold, language and run length held constant, testing after the last edit roughly doubles the odds of solving for all three models. The median failed run writes 1.6--2.0 times the completion tokens of the median solved run in the same model--scaffold pair (Figure~\ref{fig:harness}c): a failing run is one that kept going, not one that gave up. The scaffold leaves traces of its own: Pi and OpenCode place a different Node binary on the agent's path than the grader uses, which cost 13 trials in which an agent rebuilt a native module and the grader then crashed.

\section{Discussion}
\label{sec:discussion}

\paragraph{Usefulness of V2 near saturation.}
Frontier systems resolve nearly all V2 tasks, but the benchmark remains useful for four reasons. The saturation result is itself informative: on a correctly graded split the strongest systems fail fewer than 1\% of tasks, so small differences among them are not capability differences. Most deployed systems are not at the frontier: open-weight, smaller and cheaper systems span 87.2--97.5\% on the full set and 52.9--88.2\% on Hard (Table~\ref{tab:main}). This clean set supports comparisons beyond the pass rate, such as cost, time to solution and scaffold choice, which are meaningful only when every task is solvable and correctly graded. And a locked, fully solvable suite is a regression test for agent products: a failure on a task that all recent systems solve warrants investigation.

\paragraph{Lessons for benchmark maintainers.}
(1)~Use strong agents as auditors: the failure audit flagged 39 tasks the artifact audit missed, because some defects appear only when a capable agent writes a reasonable solution and the tests reject it. (2)~Repair instructions, keep tests fixed, and validate repairs blind: weakening a test to fit an instruction changes what the benchmark measures, and a blind re-implementation checks that the edited instruction suffices. (3)~Grade in a clean container: agents that cannot fetch a dependency sometimes forge one. (4)~Version the benchmark and release all grades, since new defects surface as agents improve. (5)~Check that common scaffolds run on every image: two of the scaffolds in \S\ref{sec:harness} failed to install on base images until we patched them.

\section{Conclusion}
As agents became more capable, two kinds of issues in SWE-Bench Pro became visible: tasks whose grade could disagree with the correctness of the solution, and an evaluation setup that agents could circumvent. SWE-Bench Pro V2 addresses both with a pipeline of agentic audits, engineer adjudication, blind re-implementation and verifier hardening that repaired or retired 480 of 731 tasks, and a locked protocol that grades every patch in a clean container. Frontier systems now solve almost every task, the Hard split separates them, and the full set remains a reliable benchmark below the frontier. We release the tasks and protocol, and expect the pipeline and lessons to apply to other benchmarks as agents approach their ceilings.

\section*{Limitations}
\label{sec:limitations}

\paragraph{Memorization.} The public repositories and their fixes predate the evaluated models. The protocol stops retrieval at test time, not memorization during training.

\paragraph{Scope of re-grading.} A clean-container re-grade rejects changes made outside the patch, but forgeries inside it, such as an imitation of a missing library, are left to the patch audit. 

\section*{Ethics Statement}
The tasks come from open-source repositories under their original licenses. The engineers who reviewed and repaired tasks were contracted, and no personal data about them is recorded or released. We document the exploits we observed because benchmark operators need this information to defend against them, and the countermeasures are included with the benchmark.

\bibliographystyle{unsrtnat}
\bibliography{refs}

\begin{thebibliography}{32}
\providecommand{\natexlab}[1]{#1}
\providecommand{\url}[1]{\texttt{#1}}
\expandafter\ifx\csname urlstyle\endcsname\relax
  \providecommand{\doi}[1]{doi: #1}\else
  \providecommand{\doi}{doi: \begingroup \urlstyle{rm}\Url}\fi

\bibitem[Jimenez et~al.(2024)Jimenez, Yang, Wettig, Yao, Pei, Press, and
  Narasimhan]{jimenez2024swebench}
Carlos~E. Jimenez, John Yang, Alexander Wettig, Shunyu Yao, Kexin Pei, Ofir
  Press, and Karthik Narasimhan.
\newblock {SWE}-bench: Can language models resolve real-world {GitHub} issues?
\newblock In \emph{Proceedings of the International Conference on Learning
  Representations (ICLR)}, 2024.
\newblock URL \url{https://arxiv.org/abs/2310.06770}.

\bibitem[{OpenAI}(2024)]{openai2024verified}
{OpenAI}.
\newblock Introducing {SWE}-bench {Verified}.
\newblock OpenAI blog,
  \url{https://openai.com/index/introducing-swe-bench-verified/}, 2024.

\bibitem[{OpenAI}(2026{\natexlab{a}})]{openai2026swebenchverified}
{OpenAI}.
\newblock Why {SWE}-bench {Verified} no longer measures frontier coding
  capabilities.
\newblock OpenAI blog,
  \url{https://openai.com/index/why-we-no-longer-evaluate-swe-bench-verified/},
  2026{\natexlab{a}}.

\bibitem[Deng et~al.(2026)Deng, Da, Pan, He, Ide, Garg, Lauffer, Park,
  et~al.]{deng2025swebenchpro}
Xiang Deng, Jeff Da, Edwin Pan, Yannis~Yiming He, Charles Ide, Kanak Garg,
  Niklas Lauffer, Andrew Park, et~al.
\newblock {SWE-Bench Pro}: Can {AI} agents solve long-horizon software
  engineering tasks?
\newblock In \emph{Proceedings of the 43rd International Conference on Machine
  Learning (ICML)}, volume 306 of \emph{Proceedings of Machine Learning
  Research}, 2026.
\newblock URL \url{https://arxiv.org/abs/2509.16941}.

\bibitem[Nadgir et~al.(2026)Nadgir, Kapoor, Liu, Kirgis, Orona, Rabanser,
  Bayer, Shetty, Ling, Chan-Sew, Nakagawa, Utpala, Siegel, and
  Narayanan]{nadgir2026life}
Nitya Nadgir, Sayash Kapoor, Kangheng Liu, Peter Kirgis, Matilda Orona, Stephan
  Rabanser, Tilman Bayer, Abhishek Shetty, Yue Ling, Derrick Chan-Sew, Rumi
  Nakagawa, Saiteja Utpala, Zachary~S. Siegel, and Arvind Narayanan.
\newblock Life after benchmark saturation: A case study of {CORE-Bench}.
\newblock \emph{arXiv preprint arXiv:2606.26158}, 2026.
\newblock URL \url{https://arxiv.org/abs/2606.26158}.

\bibitem[Zan et~al.(2025)Zan, Huang, Liu, Chen, Zhang, Xin, Chen, Liu,
  et~al.]{zan2025multiswebb}
Daoguang Zan, Zhirong Huang, Wei Liu, Hanwu Chen, Linhao Zhang, Shulin Xin,
  Lu~Chen, Qi~Liu, et~al.
\newblock Multi-{SWE}-bench: A multilingual benchmark for issue resolving.
\newblock In \emph{Advances in Neural Information Processing Systems (NeurIPS),
  Datasets and Benchmarks Track}, 2025.
\newblock URL \url{https://arxiv.org/abs/2504.02605}.

\bibitem[Zhang et~al.(2025)Zhang, He, Zhang, Kang, Li, Xie, Wang, Wang, Huang,
  Fu, Nallipogu, Lin, Dang, Rajmohan, and Zhang]{swebenchlive2025}
Linghao Zhang, Shilin He, Chaoyun Zhang, Yu~Kang, Bowen Li, Chengxing Xie,
  Junhao Wang, Maoquan Wang, Yufan Huang, Shengyu Fu, Elsie Nallipogu, Qingwei
  Lin, Yingnong Dang, Saravan Rajmohan, and Dongmei Zhang.
\newblock {SWE}-bench goes live!
\newblock \emph{arXiv preprint arXiv:2505.23419}, 2025.
\newblock URL \url{https://arxiv.org/abs/2505.23419}.

\bibitem[Badertdinov et~al.(2025)Badertdinov, Golubev, Nekrashevich, Shevtsov,
  Karasik, Andriushchenko, Trofimova, Litvintseva, and
  Yangel]{badertdinov2025rebench}
Ibragim Badertdinov, Alexander Golubev, Maksim Nekrashevich, Anton Shevtsov,
  Simon Karasik, Andrei Andriushchenko, Maria Trofimova, Daria Litvintseva, and
  Boris Yangel.
\newblock {SWE-rebench}: An automated pipeline for task collection and
  decontaminated evaluation of software engineering agents.
\newblock In \emph{Advances in Neural Information Processing Systems (NeurIPS),
  Datasets and Benchmarks Track}, 2025.
\newblock URL \url{https://arxiv.org/abs/2505.20411}.

\bibitem[Miserendino et~al.(2025)Miserendino, Wang, Patwardhan, and
  Heidecke]{miserendino2025swelancer}
Samuel Miserendino, Michele Wang, Tejal Patwardhan, and Johannes Heidecke.
\newblock {SWE-Lancer}: Can frontier {LLMs} earn \$1 million from real-world
  freelance software engineering?
\newblock \emph{arXiv preprint arXiv:2502.12115}, 2025.
\newblock URL \url{https://arxiv.org/abs/2502.12115}.

\bibitem[{Harbor Framework Team}(2026)]{harbor2026}
{Harbor Framework Team}.
\newblock Harbor: A framework for evaluating and optimizing agents and models
  in container environments.
\newblock \url{https://github.com/harbor-framework/harbor}, 2026.

\bibitem[Merrill et~al.(2026)]{merrill2026terminalbench}
Mike~A. Merrill et~al.
\newblock {Terminal-Bench}: Benchmarking agents on hard, realistic tasks in
  command line interfaces.
\newblock In \emph{International Conference on Learning Representations
  (ICLR)}, 2026.
\newblock URL \url{https://arxiv.org/abs/2601.11868}.

\bibitem[Pan et~al.(2024)Pan, Wang, Neubig, Jaitly, Ji, Suhr, and
  Zhang]{pan2024swegym}
Jiayi Pan, Xingyao Wang, Graham Neubig, Navdeep Jaitly, Heng Ji, Alane Suhr,
  and Yizhe Zhang.
\newblock Training software engineering agents and verifiers with {SWE-Gym}.
\newblock \emph{arXiv preprint arXiv:2412.21139}, 2024.
\newblock URL \url{https://arxiv.org/abs/2412.21139}.

\bibitem[Yang et~al.(2025)Yang, Lieret, Jimenez, Wettig, Khandpur, Zhang, Hui,
  Press, Schmidt, and Yang]{yang2025swesmith}
John Yang, Kilian Lieret, Carlos~E. Jimenez, Alexander Wettig, Kabir Khandpur,
  Yanzhe Zhang, Binyuan Hui, Ofir Press, Ludwig Schmidt, and Diyi Yang.
\newblock {SWE-smith}: Scaling data for software engineering agents.
\newblock \emph{arXiv preprint arXiv:2504.21798}, 2025.
\newblock URL \url{https://arxiv.org/abs/2504.21798}.

\bibitem[Aleithan et~al.(2024)Aleithan, Xue, Mohajer, Nnorom, Uddin, and
  Wang]{aleithan2024swebenchplus}
Reem Aleithan, Haoran Xue, Mohammad~Mahdi Mohajer, Elijah Nnorom, Gias Uddin,
  and Song Wang.
\newblock {SWE-Bench+}: Enhanced coding benchmark for {LLMs}.
\newblock \emph{arXiv preprint arXiv:2410.06992}, 2024.
\newblock URL \url{https://arxiv.org/abs/2410.06992}.

\bibitem[Wang et~al.(2026)Wang, Pradel, and Liu]{wang2025solvedissues}
You Wang, Michael Pradel, and Zhongxin Liu.
\newblock Are ``solved issues'' in {SWE}-bench really solved correctly? {An}
  empirical study.
\newblock In \emph{Proceedings of the IEEE/ACM 48th International Conference on
  Software Engineering (ICSE)}, 2026.
\newblock \doi{10.1145/3744916.3764576}.
\newblock URL \url{https://arxiv.org/abs/2503.15223}.

\bibitem[Yu et~al.(2025)Yu, Zhu, He, and Kang]{utboost2025}
Boxi Yu, Yuxuan Zhu, Pinjia He, and Daniel Kang.
\newblock {UTBoost}: Rigorous evaluation of coding agents on {SWE-Bench}.
\newblock In \emph{Proceedings of the 63rd Annual Meeting of the Association
  for Computational Linguistics (Volume 1: Long Papers)}, pages 3762--3774,
  2025.
\newblock URL \url{https://arxiv.org/abs/2506.09289}.

\bibitem[Zhu et~al.(2025)Zhu, Jin, Pruksachatkun, et~al.]{zhu2025abc}
Yuxuan Zhu, Tengjun Jin, Yada Pruksachatkun, et~al.
\newblock Establishing best practices for building rigorous agentic benchmarks.
\newblock \emph{arXiv preprint arXiv:2507.02825}, 2025.
\newblock URL \url{https://arxiv.org/abs/2507.02825}.

\bibitem[Zhong et~al.(2025)Zhong, Raghunathan, and
  Carlini]{zhong2025impossiblebench}
Ziqian Zhong, Aditi Raghunathan, and Nicholas Carlini.
\newblock {ImpossibleBench}: Measuring {LLMs}' propensity of exploiting test
  cases.
\newblock \emph{arXiv preprint arXiv:2510.20270}, 2025.
\newblock URL \url{https://arxiv.org/abs/2510.20270}.

\bibitem[Gabor et~al.(2025)Gabor, Lynch, and Rosenfeld]{gabor2025evilgenie}
Jonathan Gabor, Jayson Lynch, and Jonathan Rosenfeld.
\newblock {EvilGenie}: A reward hacking benchmark.
\newblock \emph{arXiv preprint arXiv:2511.21654}, 2025.
\newblock URL \url{https://arxiv.org/abs/2511.21654}.

\bibitem[Zhao et~al.(2026)Zhao, Srikanth, Wu, and Jiang]{zhao2026specbench}
Bingchen Zhao, Dhruv Srikanth, Yuxiang Wu, and Zhengyao Jiang.
\newblock {SpecBench}: Measuring reward hacking in long-horizon coding agents.
\newblock \emph{arXiv preprint arXiv:2605.21384}, 2026.
\newblock URL \url{https://arxiv.org/abs/2605.21384}.

\bibitem[Thaman(2026)]{thaman2026rhb}
Kunvar Thaman.
\newblock Reward hacking benchmark: Measuring exploits in {LLM} agents with
  tool use.
\newblock In \emph{Proceedings of the 43rd International Conference on Machine
  Learning (ICML)}, volume 306 of \emph{Proceedings of Machine Learning
  Research}, pages 120372--120389, 2026.
\newblock URL \url{https://arxiv.org/abs/2605.02964}.

\bibitem[Denison et~al.(2024)Denison, MacDiarmid, Barez, Duvenaud, Kravec,
  Marks, Schiefer, Soklaski, Tamkin, Kaplan, Shlegeris, Bowman, Perez, and
  Hubinger]{denison2024sycophancy}
Carson Denison, Monte MacDiarmid, Fazl Barez, David Duvenaud, Shauna Kravec,
  Samuel Marks, Nicholas Schiefer, Ryan Soklaski, Alex Tamkin, Jared Kaplan,
  Buck Shlegeris, Samuel~R. Bowman, Ethan Perez, and Evan Hubinger.
\newblock Sycophancy to subterfuge: Investigating reward-tampering in large
  language models.
\newblock \emph{arXiv preprint arXiv:2406.10162}, 2024.
\newblock URL \url{https://arxiv.org/abs/2406.10162}.

\bibitem[Baker et~al.(2025)Baker, Huizinga, Gao, Dou, Guan, Madry, Zaremba,
  Pachocki, and Farhi]{baker2025monitoring}
Bowen Baker, Joost Huizinga, Leo Gao, Zehao Dou, Melody~Y. Guan, Aleksander
  Madry, Wojciech Zaremba, Jakub Pachocki, and David Farhi.
\newblock Monitoring reasoning models for misbehavior and the risks of
  promoting obfuscation.
\newblock \emph{arXiv preprint arXiv:2503.11926}, 2025.
\newblock URL \url{https://arxiv.org/abs/2503.11926}.

\bibitem[{jacobkahn}(2025)]{kahn2025repostate}
{jacobkahn}.
\newblock Repo state loopholes during agentic evaluation.
\newblock SWE-bench GitHub issue 465,
  \url{https://github.com/SWE-bench/SWE-bench/issues/465}, 2025.

\bibitem[{joe-needham}(2025)]{needham2025prostate}
{joe-needham}.
\newblock Repo state leakage.
\newblock SWE-Bench Pro GitHub repository, issue 7,
  \url{https://github.com/scaleapi/SWE-bench_Pro-os/issues/7}, 2025.

\bibitem[{ConnorBAdams}(2026)]{adams2026gitreward}
{ConnorBAdams}.
\newblock Git reward hacking in {SWE-Bench Pro} {OSS}.
\newblock SWE-Bench Pro GitHub repository, issue 93,
  \url{https://github.com/scaleapi/SWE-bench_Pro-os/issues/93}, 2026.

\bibitem[Liang et~al.(2025)Liang, Garg, and
  Zilouchian~Moghaddam]{liang2025swebenchillusion}
Shanchao Liang, Spandan Garg, and Roshanak Zilouchian~Moghaddam.
\newblock The {SWE}-bench illusion: When state-of-the-art {LLM}s remember
  instead of reason.
\newblock \emph{arXiv preprint arXiv:2506.12286}, 2025.
\newblock URL \url{https://arxiv.org/abs/2506.12286}.

\bibitem[Ramos et~al.(2024)Ramos, Mamede, Jain, Canelas, Gamboa, and
  Le~Goues]{ramos2024memorization}
Daniel Ramos, Claudia Mamede, Kush Jain, Paulo Canelas, Catarina Gamboa, and
  Claire Le~Goues.
\newblock Are large language models memorizing bug benchmarks?
\newblock \emph{arXiv preprint arXiv:2411.13323}, 2024.
\newblock URL \url{https://arxiv.org/abs/2411.13323}.

\bibitem[Zheng et~al.(2026)Zheng, Shang, Jiang, Tian, Zhu, Ma, Yuan, and
  Zhang]{swebenchproverified2026}
Pujun Zheng, Zixin Shang, Shufan Jiang, Wenhui Tian, Dongsheng Zhu, Zerun Ma,
  Dingbo Yuan, and Qi~Zhang.
\newblock {SWE-Bench Pro Verified}: A reliable benchmark for software
  engineering agents.
\newblock \emph{arXiv preprint arXiv:2609.08149}, 2026.
\newblock URL \url{https://arxiv.org/abs/2609.08149}.

\bibitem[{OpenAI}(2026{\natexlab{b}})]{openai2026signal}
{OpenAI}.
\newblock Separating signal from noise in coding evaluations.
\newblock OpenAI blog,
  \url{https://openai.com/index/separating-signal-from-noise-coding-evaluations/},
  July 2026{\natexlab{b}}.

\bibitem[Lieret et~al.(2025)Lieret, Yang, et~al.]{minisweagent2025}
Kilian Lieret, John Yang, et~al.
\newblock mini-{SWE}-agent: The 100-line {AI} agent that solves {GitHub}
  issues.
\newblock \url{https://github.com/SWE-agent/mini-swe-agent}, 2025.

\bibitem[Yang et~al.(2024)Yang, Jimenez, Wettig, Lieret, Yao, Narasimhan, and
  Press]{yang2024sweagent}
John Yang, Carlos~E. Jimenez, Alexander Wettig, Kilian Lieret, Shunyu Yao,
  Karthik Narasimhan, and Ofir Press.
\newblock {SWE}-agent: Agent-computer interfaces enable automated software
  engineering.
\newblock In \emph{Advances in Neural Information Processing Systems
  (NeurIPS)}, volume~37, 2024.
\newblock URL \url{https://arxiv.org/abs/2405.15793}.

\end{thebibliography}

\appendix
\clearpage
\raggedbottom

\section{The Hard Split}
\label{app:hard}

The 51 tasks come from ansible (8), openlibrary (8), flipt (7), teleport (6), webclients (5), qutebrowser (4), vuls (4), element-web (4), navidrome (3), NodeBB (1) and tutanota (1). A task qualified if at least two of five model families fail on it across runs. That gave 54 candidates and 3 were removed after expert review. Every Hard task passes its reference solution, fails with an empty patch, and is solved by at least one of our runs. Of the 51 instructions, 39 were edited in the final review after selection. 

\section{Harness Study}
\label{app:harness}

\paragraph{Setup.} GLM-5.3 (thinking on), Kimi K3 (reasoning effort max) and Inkling (xhigh) ran with a 1M-token context window and no cap on thinking tokens or turns; the only cap besides the 50-minute budget is mini-swe-agent's default 30\,s per-command timeout, which is part of that scaffold. Versions: Harbor 0.22, mini-swe-agent 2.4.6, Pi 0.85.0, OpenCode 1.18.29. Before the runs we patched the task images on which Pi or OpenCode would not install (older Alpine images lacking a recent \texttt{libstdc++}, and Debian~11 images whose end-of-life package sources broke OpenCode's installer); the patches add system packages only, and all nine combinations ran on the same patched images. We release the 5{,}778 trajectories with per-trajectory feature files.

\paragraph{Attempt rule.} Unlike the runs of \S\ref{sec:experiments}, these were graded once, in the container the agent had used. Every task has one counted attempt per model--scaffold combination; an attempt that hit the budget is graded on the patch present at the deadline. An attempt that hit the budget or was killed by infrastructure \emph{without} a passing grade was run once more at lower concurrency and the re-run is the counted attempt; a re-run never replaced a graded pass, and the rule was applied identically to all nine combinations. Pi and OpenCode place a different Node binary on the agent's path than the grader uses; in 13 trials an agent rebuilt a native module against it and the grader then crashed on that module (Kimi K3: 5 under Pi, 2 under OpenCode; GLM-5.3: 2 and 2; Inkling: 2 under OpenCode), and those trials count as failures.

\paragraph{Outcomes.} Table~\ref{tab:pairs} gives the paired comparisons behind \S\ref{sec:harness}. Splitting the variance of solved-or-failed over the 642 tasks every combination attempted, task identity accounts for 31.7\%, the model for 2.65\%, the scaffold for 0.29\% and the model--scaffold interaction for 0.36\%; the residual 65.0\% is task-by-run variation.

\begin{table}[t]
\centering\footnotesize
\setlength{\tabcolsep}{3pt}
\resizebox{\ifdim\width>\columnwidth\columnwidth\else\width\fi}{!}{%
\begin{tabular}{llcrrr}
\toprule
Model & Pair ($A-B$) & only $A$\,/\,$B$ & Diff. & 95\% CI & $p$ \\
\midrule
GLM-5.3 & mini $-$ Pi & 15\,/\,21 & $-0.9$ & $[-2.8,+0.9]$ & 0.41 \\
 & mini $-$ OpenCode & 15\,/\,20 & $-0.8$ & $[-2.6,+1.0]$ & 0.50 \\
 & Pi $-$ OpenCode & 12\,/\,11 & $+0.2$ & $[-1.3,+1.6]$ & 1.00 \\
\midrule
Kimi K3 & mini $-$ Pi & 20\,/\,10 & $+1.6$ & $[-0.1,+3.2]$ & 0.10 \\
 & mini $-$ OpenCode & 8\,/\,9 & $-0.2$ & $[-1.4,+1.1]$ & 1.00 \\
 & Pi $-$ OpenCode & 6\,/\,17 & $-1.7$ & $[-3.2,-0.3]$ & 0.03 \\
\midrule
Inkling & mini $-$ Pi & 64\,/\,27 & $+5.8$ & $[+2.9,+8.6]$ & $<$0.001 \\
 & mini $-$ OpenCode & 23\,/\,34 & $-1.7$ & $[-4.0,+0.6]$ & 0.18 \\
 & Pi $-$ OpenCode & 18\,/\,66 & $-7.5$ & $[-10.2,-4.7]$ & $<$0.001 \\
\bottomrule
\end{tabular}}
\caption{Paired scaffold comparisons within a model on the same 642 tasks: tasks solved only under scaffold $A$ / only under $B$, the difference in percentage points with its 95\% interval, and an exact test on the discordant tasks. ``mini'' is mini-swe-agent.}
\label{tab:pairs}
\end{table}

\paragraph{Validation.} Pass rates are one attempt per task, a standard error of 0.6--1.4 points per combination. The activity shares of Figure~\ref{fig:harness}b and the test-after-edit analysis of \S\ref{sec:harness} come from a classifier that reads the first executable of each segment of a shell command; it matched a manual read on 147 of 150 sampled commands and on 97.2\% of sampled wall-clock, and its test-outcome detector matched on all 32 of the 40 sampled test outputs it labeled, leaving 8 undecidable. Step time includes model latency, and test-file edits are under-counted under mini-swe-agent, where edits arrive as heredocs.

\section{Reproducibility}
\label{app:repro}

\paragraph{Harness.} Harbor 0.22.0 on Modal 1.5.5. The network is set per phase: public during setup, only the model endpoint during the agent phase, public again for the verifier, because 32 tasks download dependencies while testing. Claude Code 2.1.278 with web tools disabled; Codex 0.155.1 with hosted web search disabled; mini-swe-agent 2.4.6 with no step or cost limit, so the time budget binds.

\paragraph{Budget and grading.} The budget is 3000\,s per task. The agent is stopped at 2940\,s and the patch present then is graded. The patch is the staged \texttt{git diff} of the working tree; it is applied to a fresh container built from the task image, with a three-way fallback, and the unchanged verifier runs. When the sandbox grade and the clean grade disagree, two more clean grades are taken and the majority of the three counts. Mean API spend per task on V2 was \$2.9 for the Claude Code runs, \$2.3 for Astra and \$3 for Gemini.

\paragraph{Release.} Task directories, the Hard split, the probe, capture and re-grade scripts, the exact harness flags, per-task grades in both modes for every run, and the harness-study trajectories.

\end{document}